\documentclass[aip,rsi,reprint,amsmath,amssymb]{revtex4-1}

\usepackage{graphicx, bm}
\usepackage{fp}
\usepackage{siunitx}
\usepackage{url}
\usepackage{hyperref}
\usepackage{blindtext}
\usepackage{color}
\usepackage{xfrac}
\newcommand{\fig}[1]{Fig.~\ref{fig:#1}}
\newcommand{\eqn}[1]{Eq.~\eqref{eqn:#1}}
\newcommand{\omegar}{\ensuremath{\omega_\mathrm{res}}}
\newcommand{\Vi}{\ensuremath{V_\mathrm{i}}}
\newcommand{\Vo}{\ensuremath{V_\mathrm{o}}}
\newcommand{\DD}{\ensuremath{\mathrm{d}_\mathrm{D}}}
\renewcommand{\S}{\ensuremath{S_{21}}}

\renewcommand{\vec}[1]{\ensuremath{\vb{#1}}}
\newcommand{\VBG}{\ensuremath{V_\mathrm{o}^\mathrm{BG}}}
\newcommand{\df}{\ensuremath{{\Delta\omega}/{2\pi}}}
\newcommand{\dw}{\ensuremath{\Delta \omega}}
\newcommand{\dH}{\ensuremath{\Delta H}}
\newcommand{\Ms}{\ensuremath{M_\mathrm{s}}}
\newcommand{\dmodw}{\ensuremath{\Delta \omega_\mathrm{mod} }}
\newcommand{\dmodH}{\ensuremath{\Delta H_\mathrm{mod} }}

\usepackage[sort&compress]{natbib}
\usepackage{physics}
\usepackage[colorinlistoftodos,prependcaption,textsize=tiny]{todonotes}

\begin{document}

\title{Note: Derivative divide, a method for the analysis of broadband ferromagnetic resonance in the frequency domain}

\author{Hannes Maier-Flaig}
\affiliation{Walther-Mei\ss ner-Institut, Bayerische Akademie der Wissenschaften, Garching, Germany}
\affiliation{Physik-Department, Technische Universit\"{a}t M\"{u}nchen, Garching, Germany}

\author{Sebastian T. B. Goennenwein}
\affiliation{Institut f\"ur Festk\"operphysik, Technische Universit\"at Dresden, Dresden, Germany.}
\affiliation{Center for Transport and Devices of Emergent Materials, Technische Universit\"at Dresden, Dresden, Germany}

\author{Ryo Ohshima}
\affiliation{Department of Electronic Science and Engineering, Kyoto University, Kyoto, Japan}

\author{Masashi Shiraishi}
\affiliation{Department of Electronic Science and Engineering, Kyoto University, Kyoto, Japan}

\author{Rudolf Gross}
\affiliation{Walther-Mei\ss ner-Institut, Bayerische Akademie der Wissenschaften, Garching, Germany}
\affiliation{Physik-Department, Technische Universit\"{a}t M\"{u}nchen, Garching, Germany}
\affiliation{Nanosystems Initiative Munich, M\"{u}nchen, Germany}

\author{Hans Huebl}
\affiliation{Walther-Mei\ss ner-Institut, Bayerische Akademie der Wissenschaften, Garching, Germany}
\affiliation{Physik-Department, Technische Universit\"{a}t M\"{u}nchen, Garching, Germany}
\affiliation{Nanosystems Initiative Munich,  M\"{u}nchen, Germany}

\author{Mathias Weiler}
\affiliation{Walther-Mei\ss ner-Institut, Bayerische Akademie der Wissenschaften, Garching, Germany}
\affiliation{Physik-Department, Technische Universit\"{a}t M\"{u}nchen, Garching, Germany}

\date{\today}

\begin{abstract}
Broadband ferromagnetic resonance (bbFMR) spectroscopy is an established experimental tool to quantify magnetic properties.
Due to frequency-dependent transmission of the microwave setup, bbFMR measurements in the frequency domain require a suitable background removal method.
Here, we present a measurement and data analysis protocol that allows to perform quantitative frequency-swept bbFMR measurements without the need for a calibration of the microwave setup.
The method, its limitations and advantages are described in detail.
Finally the method is applied to evaluate FMR spectra of a permalloy thin film. The extracted material parameters are in very good agreement with those obtained using a conventional analysis in field-space.
\end{abstract}
\maketitle

One key aspect in magnetic resonance spectroscopy is the investigation of the frequency dependence of the magnetic resonance linewidth which allows to extract the microscopic damping mechanisms and to quantify the Gilbert damping parameter.\cite{Maier-Flaig2017-YIG-damping, Urban2001,Tserkovnyak2002c,Woltersdorf2009,Shaw2012}
For this task, broadband FMR (bbFMR) with planar waveguide structures is very well suited and has been employed extensively in recent years.\cite{Maksymov2015}
In addition, bbFMR allows for the investigation of the magnetization dynamics of materials with various magnetic phases, and even allows to determine the magnetic phase diagram of Skyrmion-host materials.\cite{Schwarze2015,Onose2012}
An investigation of the dynamic properties of these materials, can only be performed in the frequency-domain as changing the external magnetic field will inevitably modify the magnetic configuration.
Conventional cavity-based FMR is not suitable for this task as it is naturally limited to a single frequency and therefore requires field-swept experiments.
Broadband FMR, on the other hand, is ideally suited as frequency dependent microwave spectroscopy can be performed in a wide frequency band for a series of fixed magnetic fields.\cite{Denysenkov2003}
One major challenge for this type of experiment is the frequency and temperature dependence of the microwave transmission of the setup.
This warrants a simple background removal procedure which is independent of microwave circuit calibration.

Here, we present a novel method for analyzing frequency swept FMR spectra based on computing the numerical magnetic field derivative and normalization by the microwave transmission.
The resulting data enables direct access to the high frequency magnetic susceptibility as a function of frequency. We demonstrate the reliability of the method using a permalloy (Py) sample.

The origin of magnetic resonance is the damped precession of the magnetization $\vec M$ in a static magnetic field $\vec{H}_0$ that is excited by the small magnetic field $\vec{h}_\mathrm{MW}$ oscillating with the frequency $\omega$. 
These dynamics are described by the (linearized) Landau-Lifshitz equation with the solution $\vec m = \chi \vec h_\mathrm{MW}$.\cite{Landau1935,Gilbert2004,Gurevich1996} Here, $\vec{m}$ is the the dynamic component of $\vec{M}$, $\vec{h}_\mathrm{MW}$ is the excitation field and the high frequency magnetic susceptibility is\cite{Schneider2007,Kalarickal2006}
\begin{equation}
  \label{eqn:polder}
  \chi\left(\omega, H_0\right) =  \frac{
      \omega_\mathrm{M} \left[\gamma \mu_0 H_0 - i \dw \right]
    }
    {
      \left[\omegar\!\left(H_0\right)\right]^2  - \omega^2 - i \omega \dw
    }.
\end{equation}
Here, $\omega_\mathrm{M} = \gamma \mu_0 \left|\vec{M}\right|$, $H_0 = \left|\vec{H_0}\right|$, $\gamma$ is the gyromagnetic ratio, $\omegar$ is the resonance frequency and $\dw$ is the full width at half maximum (FWHM) linewidth.
Note that the simple dependence of $\chi$ on $H_0$ is only given when magnetic anisotropies are neglected.

\begin{figure}
\includegraphics{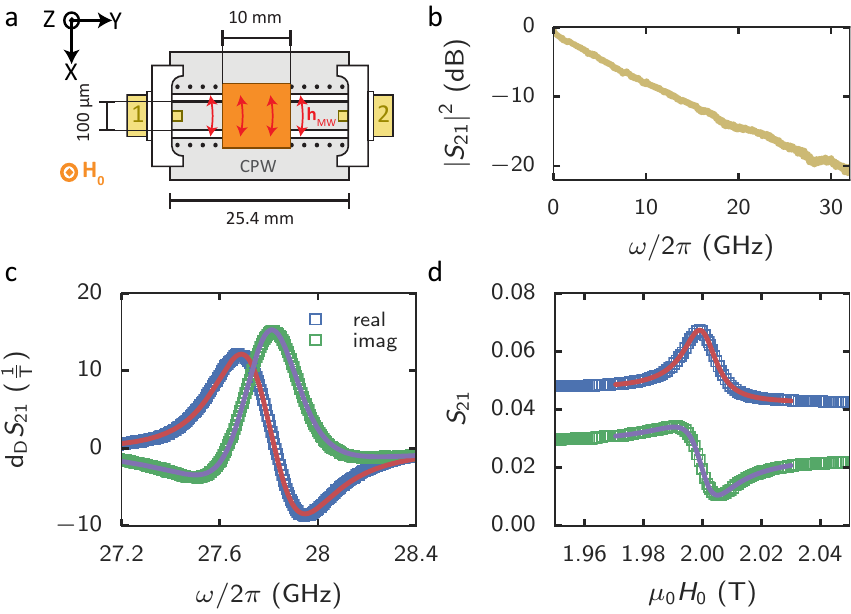}
\caption{
  \textbf{(a)} Schematic view of the CPW.
  \textbf{(b)} Raw transmission $\left|\S\right|^2$ of the setup at $\mu_0 H_0 = \SI{2}{\tesla}$.
  \textbf{(c)} $\DD\S\!\left(\omega\right)$ at $\mu_0 H_0 = \SI{2}{\tesla}$. The solid lines are a fit to $\DD \S \cdot e^{i\psi}$ with $\DD \S$ from \eqn{dd_diff} and fixed $\mu_0\dmodH = \SI{2.4}{\milli\tesla}$.
  \textbf{(d)} $\S\!\left(H_0\right)$ at $\omega/2\pi = \SI{27.8}{\giga\hertz}$. A constant offset has been subtracted for visual clarity. The solid lines are a fit to \eqn{field_fit}.
}
\label{fig:fits}
\end{figure}
The employed setup is prototypical for a bbFMR measurement setup with planar microwave structures\cite{Maier-Flaig2017-YIG-damping} (\fig{fits}~(a)).
The thin film sample is placed with the ferromagnetic layer facing the center conductor of a coplanar waveguide (CPW).
The complete CPW assembly is placed in the homogeneous  magnetic field region of an electromagnet and is connected to a vector network analyzer (VNA)\@. 
Using the VNA, the complex microwave transmission parameter of the setup $\S$ is measured as a function of $\omega$ and for a series of fixed $H_0$. 
$\S$ is given by $\S = V_\mathrm{o}/V_\mathrm{i} \cdot e^{i\phi}$ with the input voltage $\Vi$ at port~1, the measured voltage $\Vo$ at port~2 and the phase delay $\phi$. 
The FMR is detected as an voltage induced in the center conductor of the CPW by the precessing magnetization. 
The inductive voltage is given by:\cite{Silva1999}
\begin{align}
\label{eqn:coupling}
  V_\mathrm{inductive} &= 
  -i\omega A e^{i\phi} V_\mathrm{o} \chi\!\left(\omega, \vec{H_0}\right).
\end{align}
Here, $A$ is a real-valued scaling parameter that is proportional to the coupling between sample and CPW.
When the whole setup is taken into account, $V_\mathrm{inductive}$ is obscured by a complex frequency-dependent background $e^{i\phi\left(\omega\right)} \VBG\!\left(\omega\right)$ due to losses and the electrical length of the microwave setup.
In the frequency domain, $\S$ can therefore be parametrized as
\begin{equation}
  \S = \frac{-i\omega A V_\mathrm{o} \chi\!\left(\omega,H_0\right) + \VBG\!\left(\omega\right)}{V_\mathrm{i}} e^{i\phi}.
  \label{eqn:s21}
\end{equation}

The raw $\S$ data of the complete CPW assembly with sample is shown in \fig{fits}\,(b) for $\mu_0 H_0=\SI{2}{\tesla}$.
At this value of $H_0$, the FMR signal expected around \SI{28}{\giga\hertz} is barely visible as it is obscured by the frequency dependence of $\S$ of the CPW assembly. 
In principle, a suitable microwave calibration can remove this frequency dependent background.\cite{Bilzer2007}
Such a calibration is limited by the repeatability of the microwave connections and is often not sufficient to resolve the FMR signal as the background typically is temperature dependent and slowly drifting.

These issues can be resolved by using the proposed analysis method which we name \textit{derivative divide} or $\DD$.
The central differences of $\S$ with respect to $H_0$ using a finite step width $\dmodH$ are calculated and the result is divided by the central value of \S:
\begin{align}
  \DD \S &= \frac{\S\!\left(\omega,H_0 + \dmodH\right) - \S\!\left(\omega,H_0 - \dmodH\right)}{\S\left(\omega,H_0\right)\dmodH}\nonumber
  \\&\approx-i\omega A \frac{\chi\!\left(\omega, H_0 + \dmodH\right) -  \chi\!\left(\omega, H_0 - \dmodH\right)}{\dmodH}\nonumber
  \\&= -i\omega A \dv{\chi}{H_0} = -i\omega A \dv{\chi}{\omega}\frac{\partial \omega}{\partial H_0}
  =-i\omega A'\dv{\chi}{\omega}
  \label{eqn:dd}
\end{align}
To second order in the signal amplitude $A$, $\DD \S$ is proportional to the derivative of the susceptibility with respect to the external field. 
For small field steps, $\dv{\chi}{H}$ is proportional to the frequency derivative $\dv{\chi}{\omega}$ as $\chi$ varies smoothly with field and frequency.
The partial derivative $\frac{\partial \omega}{\partial H}$ therefore contributes a real-valued factor which can be absorbed in $A\rightarrow A'$.
Note that the division by $\S\!\left(\omega,H_0\right)$ is essential as it removes $\VBG$ and $\phi$ from \eqn{dd}. 
Hence, $\DD$ corrects for losses and electrical length of the setup and makes prior microwave calibration procedures obsolete.
Additionally, \textit{derivative divide} mimics a field modulation technique by calculating the derivative of the $\S$ data.
In contrast to previous work,\cite{Lo2011} \textit{derivative divide} spectra can be analyzed in frequency-space.
In principle, \textit{derivative divide} equivalent data could be obtained by combining an experimental field modulation technique with frequency-swept, calibrated VNA-FMR.
Favorably, \textit{derivative divide} evades the problems (calibration reproducibility, drift and difficult implementation) of this hypothetical experimental ansatz.

For a quantitative analysis of the $\DD\S$ spectra, one needs to account for the distortion of the line when $\dmodw$ is on the order of $\dw$.
This can be achieved by numerically calculating and fitting the central difference quotient instead of $\dv{\chi}{\omega}$ in \eqn{dd}:
\begin{equation}
  \DD \S =-i\omega A'
           \frac{\chi\!\left( \omega + \dmodw \right) - \chi\!\left( \omega - \dmodw \right)}
                {2 \dmodw}.
  \label{eqn:dd_diff}
\end{equation}
This fit formula then contains the known modulation amplitude $\dmodw = \dmodH \frac{\partial \omega}{\partial H_0} \approx  \dmodH \gamma \mu_0$.

\begin{figure}
  \includegraphics{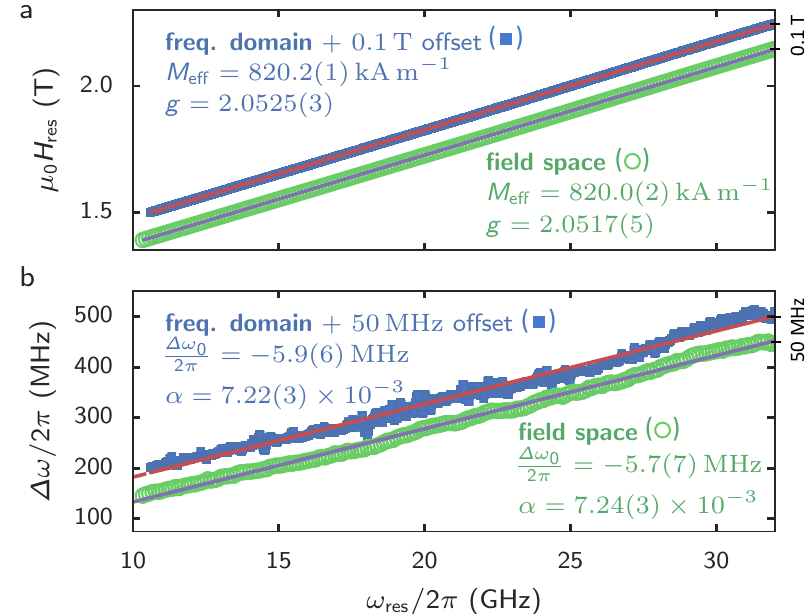}
  \caption{Results of fits of $\DD\S$ keeping $H_0$ constant [cf.\@ \fig{fits}\,(a); blue] and of a fit of $\S$ to \eqn{field_fit} keeping $\omega$ constant [cf.\@ \fig{fits}\,(b); green]. 
  \textbf{(a)} Field and frequency of resonance for the two methods (data points). Model using \eqn{kittel} (solid lines).
  \textbf{(b)} Frequency FWHM linewidth $\df$ extracted from both methods. 
  For the green data points, the field linewidth $\dH$ is converted to a frequency linewidth via $\df = \gamma \mu_0\dH$.
  For visual clarity, a vertical offset of \SI{0.1}{\tesla} resp. \SI{50}{\mega\hertz} has been added to the frequency domain results.}
  \label{fig:dispersion_damping}
\end{figure}

As an example of the use of our technique, we evaluate the well-known FMR of a \SI{30}{\nano\meter} thick Py film deposited with electron beam evaporation on a phosphorus-doped silicon on insulator substrate with lateral dimensions of $\SI{10x10}{\milli\meter}$.
$\S\!\left(\omega\right)$ is recorded with a Keysight VNA (N5244A PNA) for fixed $H_0$ in the range of $1.4$ to \SI{2.5}{\tesla} with a resolution of $\SI{0.8}{\milli\tesla}$.
\textit{Derivative divide} heavily suppresses the frequency and field dependent background signal.
Hence, automated extraction of $\omegar$ is straightforward from the field-frequency map of $\DD\S$.
We fit cuts of $\DD\S$ at fixed $H_0$ to \eqn{dd_diff} multiplied with a phase factor $e^{i\psi}$. 
In the fit, we take $\gamma$ and $\left|\vec{M}\right|$ to be constant and determine them in a subsequent fit of $\omegar\!\left(H_0\right)$.
We use the known $\mu_0\dmodH = \SI{2.4}{\milli\tesla}$ corresponding to $\dmodw = \SI{91}{\mega\hertz}$ for the fit. 
In our conducting Py sample inductively or capacitively generated currents flowing in the sample give rise to a finite phase shift $\psi$.\cite{Berger2016}
Note that for insulating ferromagnets no additional phase shift is expected ($\psi = 0$).
An example of the excellent agreement of fit and data is shown in \fig{fits}\,(c).
The fitted resonance frequency \omegar\ and linewidth \dw\ are displayed in \fig{dispersion_damping} (green data points).
Fitting $\omegar\left(H_0\right)$ to the Kittel equation for a thin ferromagnetic film magnetized perpendicular to its plane [solid lines in \fig{dispersion_damping}\,(a)]
\begin{equation}
  \omegar = \gamma \mu_0 \left(H_0 - \Ms\right)
  \label{eqn:kittel}
\end{equation}
gives a gyromagnetic ratio of $\gamma = \SI{28.727 +- 0.004}{\giga\hertz\per\tesla}$ (corresponding to $ g = \gamma \frac{\hbar}{\mu_\mathrm{B}} = 2.0525(2)$) and a saturation magnetization of $\Ms = \SI{820.1 +- 0.1}{\kilo\ampere\per\meter}$. Both fit parameters are in very good agreement with literature.\cite{Schoen2015}
The linewidth evolution is accurately described with a Gilbert-like damping model\cite{Rossing1963}
\begin{equation}
  \dw = 2 \alpha \omega + \dw_0,
  \label{eqn:gilbert}
\end{equation}
yielding a Gilbert damping parameter $\alpha = 0.00722(3)$ which is typical for Py thin films.\cite{Schoen2015} The inhomogeneous linewidth $\dw_0/2\pi = \SI{-5.9 \pm 0.6}{\mega\hertz}$ is very close to zero indicating excellent homogeneity of our film. The small negative value can be explained by uncertainties in the field readout and  deviations from linearity of $\dw(f)$.

Finally, in order to confirm that \textit{derivative divide} does not distort \dw\ or \omegar,  we perform the conventional procedure of fitting fixed frequency cuts of $\S$ to
\begin{equation}
  \S\!\left(H_0\right)|_\omega =\
   -i\omega \,A e^{i\phi}\chi\!\left(\omega,H_0\right) + B + C\cdot H_0,
  \label{eqn:field_fit}
\end{equation}
with the complex parameters $B$ and $C$ which parametrize the field dependent background of $\S$. 
The model fits the data accurately as shown for $\omega/2\pi=\SI{27.8}{\giga\hertz}$ in \fig{fits}\,(d).
The fit results are shown in \fig{dispersion_damping} (blue data points). 
The material parameters extracted using the two methods are in excellent agreement which shows that \textit{derivative divide} allows to faithfully determine $\dw$ and $\omegar$ even when $\dmodw$ is not small compared to $\dw$. 
Note that \textit{derivative divide} results in a similar signal-to-noise ratio as field-swept FMR as visible by comparing Fig.~\ref{fig:fits}\,(c) and (d). Therefore, also the sensitivity is similar to field-swept bbFMR which depends in particular on the sample shape and the center conductor width.\cite{Silva2016,Denysenkov2003}

In conclusion, we presented a novel method for analyzing broadband ferromagnetic resonance data in frequency domain without the need for a microwave circuit calibration.\cite{Bilzer2007}
The method allows to efficiently separate the complex signal induced by the precessing magnetization from a magnetic field dependent background.
The method is analogous to an experimental field modulation technique and additionally removes the frequency dependent transmission background. Thereby, it makes an analysis of ferromagnetic resonance in frequency domain possible. 
We demonstrate the method using a permalloy thin film%
 that poses a well understood exemplary material system%
.
We find excellent agreement of $\omegar$ and $\dw$ obtained by derivative divide and conventional field-swept bbFMR.
We conclude that the our method is well suited for the analysis of ferromagnetic resonance in frequency domain.
The complete set of raw data, a reference implementation of \textit{derivative divide} and automated fitting models for $\S$ are freely available at Ref.~\citenum{Maier-Flaig2017-dd-data}.

The authors thank M. S. Brandt for lending the VNA used in this work.
We gratefully acknowledge funding via the priority program 1538 Spin Caloric Transport (spinCAT) of Deutsche Forschungsgemeinschaft (projects GO 944/4, GR 1132/18), SFB 631 and the priority program 1601 (HU 1896/2-1).

\bibliography{references}
\end{document}